**T.H. Geballe [a] and M. Marezio [b]**

[a] *Department of Applied Physics and Materials Science, Stanford University, Stanford, CA 94305, USA*
[b] *CRETA/CNRS, 38042 Grenoble cedex 9, France*

**Comment on:**
**Out of plane effect on the superconductivity of $Sr_{2-x}Ba_xCuO_{3+\delta}$ with $T_c$ up to 98 K**
W.B. Gao, Q.Q. Liu, L.X. Yang, Y.Yu, F.Y. Li, C.Q. Jin and S. Uchida
Phys. Rev. B **80**, 094523 (2009) – Published September 24, 2009

We applaud the further investigation of the unexpected superconductivity in overdoped $Sr_2CuO_{3+x}$ by Gao et al., because there might be unusual physics responsible for the enhanced superconductivity that is observed. The lack of single phase material makes it difficult to interpret the data but we believe there exists relevant data in the literature that lead to different deductions than those presented by Gao et al. We suggest an alternative model that does not agree with their interpretation that the increase in cell volume due to the Ba doping is responsible for the marked increases the $T_c$ of $Sr_2CuO_{3+x}$ upon doping with Ba and that the increase in the Cu-O in-plane bond length is the dominant factor. In our model there is no significant increase in $T_c$ with Ba doping when the data are normalized. We present additional data that show the effect of disorder in suppressing $T_c$ is overestimated. We further suggest that the agreement the authors find in the refinement of their x-ray powder diffraction data with their assumed structure is an artifact of the procedure used and does not provide meaningful evidence for the assumed location of the oxygen vacancies.

There has been considerable uncertainty in identifying the phase responsible for the unusually high $T_c$ of $Sr_2CuO_{4-v}$ following its first synthesis under high pressure in the presence of $KClO_4$ by Hiroi et al. in 1993 [1]. Those authors and subsequent workers used the formula $Sr_2CuO_{3+x}$ because the synthesis involves adding oxygen to the $Sr_2CuO_3$ starting material. However, the unit cell obtained is of the $K_2NiF_4$ type so that it is more appropriately written as $Sr_2CuO_{4-v}$. The synthesis is difficult; it requires high pressures and temperatures in the presence of a strong oxidizing agent and the retention of the metastable superconducting phase in a multiphase sample. The results were confirmed and significantly extended in a series of further investigations [2, 3, 4, 5]. However, a subsequent investigation [6] that employed a much weaker oxidizing agent, $KClO_3$, than the $KClO_4$ used by the earlier workers, led to the identification of a trace amount of superconductor in a small region along one side of the sample.

This led to the suggestion [6] that all the earlier results were also spurious due to chloride contamination. A re-examination of the procedures and analysis employed on [6] shows the suggestion is without merit [7, to be published]. Nevertheless, a cessation of further research ensued until C. Q. Jin and co-workers [8] introduced an elegant synthesis that avoided the possibility of the chloride contamination and furthermore provided a method for determining the oxygen stoichiometry simply from the starting materials.

The intriguing question of what is responsible for the enhancement of $T_c$ in $Sr_2CuO_{4-v}$, a factor of 2 higher than other members of the extensively studied $RE_2CuO_4$ cuprate family, remains unanswered. Gao et al. [9] in the present work have attempted to gain insight by investigating the volume dependence of $T_c$. They synthesized a series of Ba doped compounds up to $Sr_{1.4}Ba_{0.6}CuO_{4-v}$ using Ba to increase the size of the unit cell while holding the doping concentration constant. The onset $T_c$ is 98K for the un-annealed $Sr_{1.4}Ba_{0.6}CuO_{4-v}$ sample and they attribute the increase over the 75K found for similarly un-annealed $Sr_2CuO_{4-v}$ [8] to the larger size of the Ba ion. They apparently assume that since the $T_c$ of the quenched metastable of $Sr_2CuO_{4-v}$ was increased to 95K by annealing for 12 hours at 300C a similar increase would occur if the Ba-doped sample could be annealed. We think that such an assumption is not likely to be valid and offer a simple explanation of the data as follows. When the $Sr_2CuO_{4-v}$ was annealed above 300K the superconductivity vanished along with a loss of weight indicating a decrease in the oxygen content [8]. We note therefore that 300C must be close to $T_{ms}$ where $T_{ms}$ is defined as the highest temperature for which the metastable superconducting phase can be retained. On the other hand $T_{ms}$ must be close to room temperature for $Sr_{1.4}Ba_{0.6}CuO_{4-v}$, because, quoting from [9] "its superconductivity was immediately lost upon heating even at 150C." Thus in both cases the maximum $T_c$ is achieved when $T_{ms}$ is approached. Making the reasonable assumption that the oxygen mobility increases markedly, likely exponentially, as $T_{ms}$ is approached, it follows that optimum ordering of the oxygen is achieved in short laboratory time scales at temperatures near $T_{ms}$. These considerations suggest an alternative to the explanation given in [9]. Namely, the larger Ba decreases $T_{ms}$ to near room temperature which in turn allows for optimum order without any annealing; contrary to the model suggested in [9], almost no further increase in $T_c$ is possible. When $T_c$s of the two sets of samples are normalized to $T_{ms}$ there is no significant affect that can be attributed to the change in volume; the difference between 95K and 98K is well within experimental error. When the $Sr_2CuO_{4-v}$ is cooled through $T_{ms}$ during the final stage of its synthesis it does not have sufficient time to achieve the optimum oxygen ordering.

Gao et al. [9] present x-ray diffraction data in their Fig. 5 selected to support the idea that Tc of the single-layered cuprates increases monotonically as a function of the length of the in-plane Cu-O bond with the highest $T_c$ has the longest Cu-O bond. They note that the opposing effect of disorder is minimal because the Ba and Sr ions are isovalent. [We question their estimate of the magnitude of the effect of disorder as discussed below.] The seemingly good agreement they obtain is consistent with Attfield et al.'s [10] results for other single layer cuprates. However, it is surprising to us that the values for the $Sr_2CuO_{4-v}$ from their earlier work [8], and which should provide the most direct comparison, are not included. In the figure below we have re-plotted their Fig. 5 with those points for annealed and unannealed $Sr_2CuO_{4-v}$ samples included. The added points cast doubt on the conclusion [9] that the Cu-O planar distance is the decisive parameter in determining trends in $T_c$. In addition, we have included neutron diffraction result of Shimikawa et al. [2] because it gives the most reliable structural refinement for the location of the oxygen. Even though the superconductivity is degraded we do not believe the results should be discounted for the following reason. After the neutron investigation was completed the sample was ground to a powder, a treatment designed to destroy spurious superconducting connections. The following ac susceptibility measurement then gave a rather sharp transition ~ 60K with a magnitude corresponding to a superconducting volume fraction of ~ 7%. This fraction is not insignificant for multiphase samples where considerable pinning is likely. There is obviously a need for a better sample but in our opinion the superconductivity is sufficiently robust to justify inclusion in the replot of Fig 5.

Gao et al. [9] use the Cu-O distance given by x-ray diffraction of the Ba-doped $Sr_2CuO_{4-v}$ despite the earlier TEM work with $Sr_2CuO_{4-v}$ that led to the conclusion that the superconducting phase was not the majority phase [8]. Unfortunately, they were unable to conduct electron imaging and diffraction experiments because the electron beam decomposed the metastable sample, probably because $T_{ms}$ is so near to room temperature. In any case, the various phases found when $SrO_2$ is used as the oxidizing agent are the result of different oxygen vacancy periodicities and should not affect the Cu-O in plane bond length to first order. In the earlier investigations [2, 3, 4, 5] there is no evidence for the observed superconductivity being due to a minority phase, perhaps because of the simpler reaction paths when $KClO_4$ is used as the oxidizing agent.

The model employed [9] gives a disorder parameter in $Sr_{1.4}Ba_{0.6}CuO_{4-v}$, which predicts that the $T_c$ of "disorder-free" material would be 130K or higher [9]. That model can be tested by comparing $(LaSr)_2CuO_4$ samples that have very large disorder parameters with oxygen-staged cuprates, $La_2CuO_{4+d}$ [11] for which the calculated disorder parameter is zero. In contradiction to the prediction of the model, there is only a small difference when the optimum $T_c$s in the respective series are compared and is evidence that the disorder should have little effect in degrading the $T_c$ of $Sr_{1.4}Ba_{0.6}CuO_{4-v}$

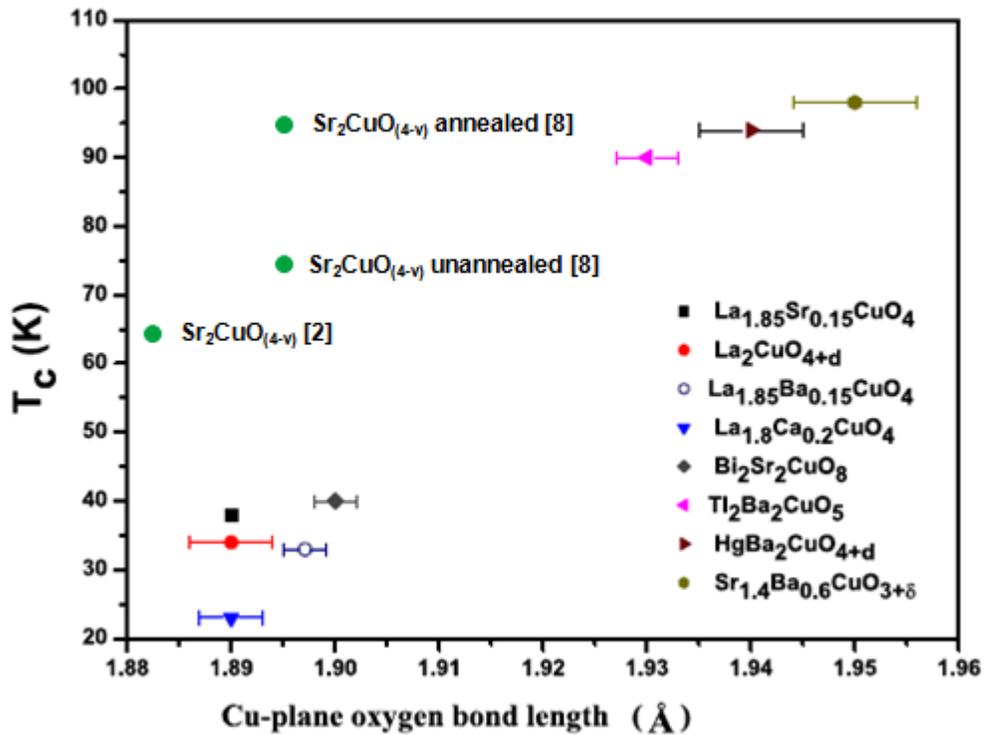

Fig. 5 of ref. 9 with and added point from ref. 8 as described in the text.

A second intriguing question is the location of the oxygen vacancies. The experimental value for the oxygen stoichiometry determined from the nominal composition [8.9] is v = 0.6. The simplest interpretation for a homogeneous distribution is that the Cu valence is +2.8. This requires the samples to be overdoped in the metastable phase well beyond the limit of + 2.3 where Fermi liquid behavior is found universally beyond the dome in doped cuprates. In [8] and [9] the assumption is made that the $CuO_2$ layers cannot contain oxygen vacancies based upon the overwhelming evidence that stoichiometric $CuO_2$ layers are a basic requirement for superconductivity. The authors, therefore, are forced to assume that the vacancies are on the apical oxygen sites in the SrO layers. They are thus forced to disregard Shimikawa's neutron data that show the vacancies are in the $CuO_2$ layers. They are able to obtain a good fit between the measured and calculated x-ray diffraction patterns (actually for a sample containing a lower concentration of Ba). However, the crucial parameters, such as the position of the apical oxygen were not refined. This position, the z parameter, was kept fixed whereas its occupation factor was varied and very surprisingly went to 0.58. This value corresponds to v = .84 and a Cu valence of 2.32 values, which would imply simple Fermi liquid metal behavior.

In conclusion, we note that Jin and his colleagues have reopened a neglected potentially rich field in new region of the phase diagram of high-temperature cuprate superconductors [8, 9]. It is not straightforward to interpret the results because the samples are multiphase. In this comment we question their interpretations and suggest other models. The evidence for $T_c$ being dominated by the Cu-O bond length in Fig 5 depends upon an arbitrary selection of data. Attributing the enhanced $T_{cs}$ that are observed to the ordering of vacancies on the apical oxygen sites is a speculation that has only questionable experimental support. Credible experimental evidence that the vacancies exist in the apical oxygen sites has not been provided and earlier results [2, 3, 4, 5] that employed a different synthesis method and give evidence for the oxygen vacancies being in the CuO2 layers are disregarded . Better samples are needed before this issue can be resolved. The possibility that there is a whole region of highly overdoped cuprates with the same structure discovered by Bednorz and Mueller 24 years ago is an enticing indication of undiscovered mechanisms and should inspire further research.

## Acknowlegments

We have enjoyed conversations with a number of colleagues, particularly S.A. Kivelson, D.S. Scalapino, and E. Berg. The work at Stanford was supported in part by the Air Force Office of Scientific Research.